\documentclass[a4paper,fleqn,usenatbib]{mnras}

\usepackage{newtxtext,newtxmath}
\usepackage[T1]{fontenc}
\usepackage{ae,aecompl}

\usepackage{graphicx}	% Including figure files
\usepackage{amsmath}	% Advanced maths commands
\usepackage{amssymb}	% Extra maths symbols

\title[Polarization and dusty environment around SNe Ia]{Polarization as a probe of dusty environments around Type Ia supernovae: radiative transfer models for SN 2012dn}

\author[T. Nagao et al.]{
Takashi Nagao,$^{1,3}$\thanks{E-mail: nagao@kusastro.kyoto-u.ac.jp}
Keiichi Maeda,$^{1}$
and Masayuki Yamanaka$^{2}$
\\
$^{1}$Depertment of astronomy, Kyoto University, Kitashirakawa-Oiwake-cho,
 Sakyo-ku, Kyoto 606-8502, Japan\\
$^{2}$Hiroshima Astrophysical Science Center, Hiroshima University, Kagamiyama, Higashi-Hiroshima 739-8526, Japan\\
$^{3}$Research Fellow of Japan Society for the Promotion of Science (DC2)
}

\date{Accepted 2018 February 22. Received 2018 February 01; in original form 2017 November 07}

\pubyear{2018}

\begin{document}
\label{firstpage}
\pagerange{\pageref{firstpage}--\pageref{lastpage}}
\maketitle

\begin{abstract}
Geometry of circumstellar (CS) medium around supernovae (SNe) provides important diagnostics to understand the nature of their progenitors. In this paper, properties of CS dust around SN 2012dn, a super-Chandrasekhar candidate Type Ia supernova (SC-SN), have been studied through detailed three dimensional radiation transfer simulations. With the detected near-infrared excess from SN 2012dn, we show that it has a disk-like dusty CS environment whose mass is roughly consistent with a branch of an accreting white dwarf system (the single degenerate scenario). We show that a similar system should produce polarization signals up to $\sim 8$ \% in the $B$ band, depending on the viewing direction if polarimetric observation is performed. We predict that the maximum polarization is reached around $\sim 60$ days after the $B$-band maximum. We show that the temporal and wavelength dependence of the polarization signals, together with other unique features, can be easily distinguished from the interstellar polarization and intrinsic SN polarization. Indeed, the small polarization degree observed for normal Type Ia SNe (SNe Ia) can constrain a parameter space in the CS dust mass and distribution. We thus encourage multi-band polarimetric observations for SNe Ia, especially for outliers including SC-SNe for which some arguments for the single degenerate scenario exist but the polarization data are very rare so far.
\end{abstract}


%%%%%%%%%%%%%%%%%%%%%%%%%%%%%%%%%%%%%%%%%%%%%%%%%%

%%%%%%%%%%%%%%%%% BODY OF PAPER %%%%%%%%%%%%%%%%%%
\section{Introduction}
Type Ia supernovae (SNe Ia) have been used as accurate cosmological distance indicators \citep[e.g.,][]{Riess1998, Perlmutter1999}. However, the progenitors of SNe Ia are still under active debate, though some progenitor scenarios have been proposed for several decades. The proposed scenarios are basically classified into the following three classes: the so-called single-degenerate (SD), double-degenerate (DD), and core-degenerate (CD) scenarios \citep[e.g.,][]{Whelan1973, Sparks1974, Nomoto1982, Iben1984, Webbink1984, Livio2003}. Recently, various kinds of SN Ia outliers have been observationally reported (e.g., so-called super-Chandrasekhar candidate SNe Ia; hereafter SC-SNe). It is important to understand relations between different types of SNe Ia and the progenitor scenarios \citep[see][for a review]{Maeda2016b}. 

The properties of circumstellar medium (CSM) around SNe Ia provide powerful diagnostics to reveal the nature of the progenitors. Studying the narrow absorption systems toward SNe suggests that a fraction of SNe Ia, including a normal class, could have a substantial amount of CSM \citep[e.g.,][]{Patat2007, Simon2009, Sternberg2011, Foley2012, Maguire2013, Sternberg2014}, while attributing the observed absorption systems either to CSM or interstellar medium is challenging \citep[][for SN 2014J observations]{Maeda2016a}. Radio and X-ray emissions from the SN-CSM interaction have never been detected for normal SNe Ia, placing especially strong constraints on the existence of CSM around SNe 2011fe and 2014J within $\sim 0.01$ pc \citep[][]{Margutti2014, Perez-Torres2014}. The information on CSM at a larger scale can be obtained by searching for near-infrared (NIR) emissions through the CS echo, and upper limits have been placed on the CSM mass for normal SNe \citep[][]{Maeda2015}.

In the method to use the NIR echo, observed NIR emission from an SN with circumstellar (CS) dust consists of two components from both CS dust and an SN itself \citep[e.g.,][, hereafter Paper I]{Maeda2015, Nagao2017a}. The CS dust echo component, which tells us about the CS environment around the SN, is estimated by subtracting intrinsic NIR emission from the SN in the observed NIR light curves. This process requires to accurately estimate the intrinsic NIR light curves of SNe. However, there are no established templates of the late-time NIR light curves for peculiar types of SNe Ia. In addition, the intrinsic NIR emission could include radiation not only from SNe themselves but also from newly formed dust \citep[e.g.,][for the dust formation in SNe Ia]{Nozawa2011}, whose contribution might be different for different SNe. Even in a case where the pure CS dust component would be robustly derived, there can still be degeneracies in deriving the geometry of CS dust (Paper I).

A polarimetric observation is potentially a powerful and alternative tool to reveal the CS environment around SNe \citep[e.g.,][]{Wang1996, Mauerhan2017, Nagao2017b}. Even though the emissions from SNe Ia themselves are not highly polarized \citep[see][for a review]{Wang2008}, scattering of the SN light by aspherically-distributed CS dust can produce a net polarization.

A NIR excess that originates from CS dust was recently discovered for an SC-SN 2012dn \citep{Yamanaka2016}. The NIR excess appeared at $\sim30$ days after the $B$-band maximum, then increased to the peak at $\sim60$ days. In Paper I, properties of the CS dust around SC-SN 2012dn have been investigated though calculations of NIR echoes from CS dust, though using a simple method without taking into account extinction effects for an input SN light. They derived the CSM dust mass of $7.0 \times 10^{-4} M_{\odot}$, which corresponds to the (gas) mass loss rate of $1.2 \times 10^{-5} M_{\odot} \rm{yr}^{-1}$ assuming the gas-to-dust ratio of 100. Further, the delayed emergence of the NIR excess readily rejects the spherically symmetric CSM distribution, reaching to the conclusion that the favored conflagration is either a disk-like or blob/jet-like. The spectral energy distribution (SED) is fully consistent with the carbon dust composition with the temperature expected from the echo model, while the required temperature of the CS dust is too high to be silicate. In \citet{Yamanaka2016} and Paper I, the (scaled) NIR light curves of another SC-SN 2009dc are assumed to represent the intrinsic SN light curves of SC-SN 2012, because its NIR light curves show very similar shapes to those of SC-SN 2012dn until $\sim 30$ days from the $B$-band maximum.

In this paper, we examine a detailed distribution of the CS dust around SC-SN 2012dn using a detailed radiation transfer code. First, we verify the validity of the distribution of the CS dust studied in Paper I, which was derived using a simple method to calculate the NIR echoes ignoring extinction effects for an input SN light. Based on the refined estimation on the amount and geometry of the CS dust, we then calculate time evolution of polarization expected from such dusty CS environment. In Section 2, we summarize our methods. In Section 3, we present our CS dust model. In Section 4, we describe our results. The paper concludes in Section 5 with discussions.

\section{Method}
We perform three-dimensional (3D) radiation transfer calculations to study NIR echoes and polarized-scattered echoes from CS dust around SNe. For this purpose, we use the 3D Monte Carlo radiation transfer code developed by \citet{Nagao2016} and \citet{Nagao2017b}. This time-dependent transfer code can treat absorption and scattering processes, on an input SN light, by CS dust. The static CS dust with an arbitrary spatial distribution can be mapped on our simulation grids. The size of the SN ejecta is negligible as compared to the spatial extent of the CSM, and thus the source of the input SN light is assumed to be a point source located at the origin. As in Paper I, we determine temperature of CS dust ($T(t,r)$), assuming radiative equilibrium as follows:
\begin{equation}
\int^{\infty}_{0} \frac{L_{\rm{SN},\nu}(t)}{4 \pi r^2} \kappa_{\rm{abs},\nu} {\rm d}\nu = 4 \pi \int^{\infty}_{0} \kappa_{\rm{abs},\nu} B_{\nu} (T(t,r)) {\rm d}\nu,
\end{equation}
where $L_{\rm{SN},\nu}(t)$ is an incoming SN luminosity that is emitted from the SN at time $t$, $B_{\nu}(T)$ is the Planck function with temperature $T$, and $\kappa_{\rm{abs}, \nu}$ is a mass absorption coefficient of the dust at frequency $\nu$. This treatment for the temperature is a good approximation for the optically thin limit, and this is applicable to the CS dust studied here (see \S 4.1). In calculating polarization due to dust scattering, we use a widely adopted method following \citet{Code1995}, which was incorporated in our simulation code by \citet{Nagao2017b}. At each scattering, the original values of Stokes parameters of a photon are converted into the new values using the scattering matrix of \citet{White1979}. For the peak linear polarization in the matrix, we assume $p_{l} = 0.5$, which is widely used in other studies \citep[e.g.,][]{Code1995}.

As an input SN light, we adopt Hsiao's template spectral evolution \citep{Hsiao2007}, which is the same setup as adopted in Paper I. This template spectral series, as created by averaging the spectral series of a number of normal SNe Ia, covers the spectral evolution from early to late phases from the UV through NIR bandpasses. The light curves of SC-SN 2012dn in the optical bands are very similar to those in the Hsiao's template, despite its spectral identification as an SC-SN \citep[e.g.,][]{Yamanaka2016}.

For optical properties of the CS dust, we adopt the same graphite dust model with that adopted in Paper I (see Section 3.1 in Paper I). In the dust model, the shape of the dust grains is assumed to be spherical. We adopt the so-called MRN size distribution, $\propto x^{-3.5}$, where $x$ is a radius of a dust grain \citep[][]{Mathis1977}. The maximum and minimum radii of the dust grains are set to be 0.2 and 0.05 $\mu$m, respectively. Using these values, the values of the mass absorption and scattering coefficient and the scattering asymmetry parameter are calculated using Mie theory. 

We note that the dust composition is strongly constrained for the CS dust around SC-SN 2012dn through the NIR emission: carbon dust grains are favored. In \citet{Yamanaka2016}, they fit the observed SED of the NIR excess testing both amorphous carbon and astronomical silicate dust, and found that the temperature to explain the SED with silicate dust is significantly higher than the evaporation temperature of astronomical silicate, $\sim 1000$K \citep[][]{Nozawa2003}.

\section{Models}
In Paper I, the NIR excess of our target, SC-SN 2012dn, has been explained by the NIR echo from the CS dusty disk or jet, while a spherically distributed CS dust could not explain the delayed emergence of the NIR excess as observed for SN 2012dn \citep[Section 1; see][]{Yamanaka2016}. In this study, we adopt the same geometry of the CS dust distribution with those adopted in Paper I (the disk and jet models, see Section 3.2 and Figure 2 in Paper I). 

The parametrization of the disk and jet models is the same as in Paper I. The disk model is specified by the inner and outer radii ($r_{\rm{in}}$ and $r_{\rm{out}}$, respectively) and the opening angle ($\theta_{0}$). The observables are dependent on the viewing direction, which is denoted as $\theta_{\rm{obs}}$ (the angle between the observer and the axis of the rotational symmetry, i.e., the polar direction of the disk). The description of the bipolar jet model follows the same terminology. In both models, the radial density distribution of the CS dust is assumed to follow $\rho_{\rm{dust}}(r) = \rho_{\rm{dust}}(r_{\rm{in}}) (r/r_{\rm{in}})^{-2}$, as expected from a stationary mass loss from a progenitor system. In summary, we have $\theta_{0}$, $r_{\rm{in}}$, $r_{\mathrm{out}}$, $\theta_{\rm{obs}}$, and $\rho_{\rm{dust}}(r_{\rm{in}})$ as our tunable parameters for both configurations.

\section{Results}
\subsection{Revisiting the NIR excess in SC-SN 2012dn with transfer calculations}
In this subsection, we reinvestigate the NIR excess in SC-SN 2012dn as a thermal echo from CS dust. In Paper I, the best-fit parameters for the CS dust to explain the NIR excess in SC-SN 2012dn have been derived using a simple method, without taking into account effects of extinction on the input SN light. The best-fit parameters derived in Paper I are shown in Table 1. In this subsection, we revisit the same issue but using the 3D Monte Carlo radiation transfer code, taking into account the extinction effects in a self-consistent manner. 
\begin{table*}
\caption{The best-fit parameters in Paper I}
\begin{tabular}{ccccccccc}
\hline
model & $\theta_{0}$[degree] & $r_{\rm{in}}$[pc] & $r_{\rm{out}}$[pc] & $\theta_{\rm{obs}}$[degree] & $\rho_{\rm{dust}}(r_{\rm{in}})$ [g/cm$^{3}$] & dM/dt(gas+dust)[M$_{\bigodot}$/yr] & $\tau_{\parallel}$(B) & $\tau_{\perp}$(B)\\
\hline \hline
disk & 20 & 0.04 & 0.1 & 0 & 2.4E-22 & 1.2E-5 & 1.303 & 0.383\\ \hline
jet & 20 & 0.04 & 0.1 & 90 & 2.7E-21 & 1.2E-5 & 14.897 & 4.378\\ 
\hline
\end{tabular}
\begin{minipage}{.88\hsize}
Notes. In Paper I, the values of $\rho_{\rm{dust}}(r_{\rm{in}})$ and $\dot{\rm{M}}$ for the jet model were wrong. The corrected values are shown here.
\end{minipage}
\end{table*}
\begin{table*}
\caption{Best-fit parameters in this study}
\begin{tabular}{ccccccc}
\hline
model & $\theta_{0}$[degree] & $r_{\mathrm{in}}$[pc] & $r_{\mathrm{out}}$[pc] & $\theta_{\mathrm{obs}}$[degree] & $\rho_{\mathrm{dust}}(r_{\mathrm{in}})$ [g/cm$^{3}$] & dM/dt(gas+dust)[M$_{\bigodot}$/yr]\\
\hline \hline
disk & 50 & 0.04 & 0.1 & 0 & 1.3E-22 & 1.6E-5\\ 
\hline
\end{tabular}
\end{table*}

Figure 1 shows the updated light curves of the NIR echoes computed with the transfer code, for the `best-fit' models in Paper 1. For comparison, the original light curves for the same disk model, as computed in Paper I with the simplified method, are also shown. The revised light curves in the disk model reproduce those derived in Paper I reasonably well. In the late phase, they are slightly fainter than those derived in Paper I, because of the effects of the extinction by the CS dust for an input SN light. The NIR emission emerges earlier in the present simulation than in Paper I, since the latter overestimated the extinction on the echo emission due to the assumed uniform extinction in space. For the jet model, the derived light curves are substantially fainter than those derived in Paper I, again because of the effects of the extinction (see below for details). 
\begin{figure}
  \includegraphics[width=\columnwidth]{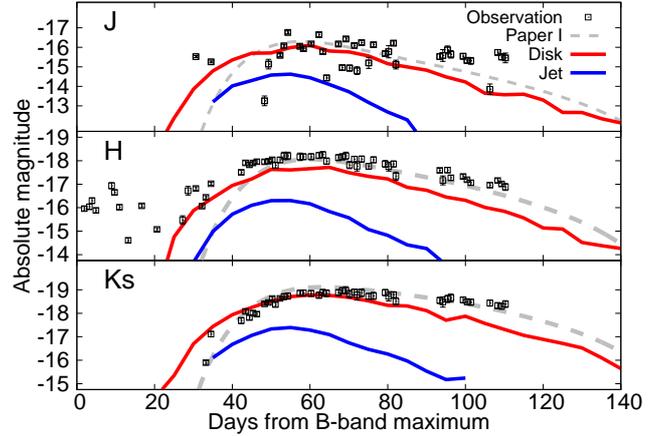}
  \caption{$J$-, $H$-, and $K_s$-band light curves in the best-fit disk and jet models of Paper I calculated using the Monte Carlo code (red solid lines for the disk model, magenta solid lines for the jet model). The gray dashed lines show the original light curves in the disk model of Paper I calculated using the simple method. The light curves of the NIR excess seen in SC-SN 2012dn are also shown (black squares).}
\end{figure}

With the detailed radiation transfer simulations in this paper, we derive the amount and distribution of the CS dust around SC-SN 2012dn more precisely than in Paper I. Under our parameterization on the CS dust distribution, there are four parameters: $\theta_0, r_{\rm{in}}, r_{\rm{out}},\rho_{\rm{dust}}(r_{\rm{in}})$. For $\theta_{\rm{obs}} \sim 0$ degree, which is the case for the CS dusty disk around SC-SN 2012dn, the result is not sensitive to $r_{\rm{in}}$ and $r_{\rm{out}}$ (see Paper I). Thus, we adopt the same values of $r_{\mathrm{in}}$ and $r_{\mathrm{out}}$ as in Paper I. Figure 2 shows the light curves of the NIR echo for the new best-fit disk model, which is derived by calculating NIR echoes with various values of $\theta_{0}$ (10, 20, 30, 40, 50, 60, 70, 80 and 90 degree) toward $\theta_{\rm{obs}}=0$ degree. The best-fit model is judged by visual inspection. The updated best-fit parameters are shown in Table 2. In the jet model, we find no solution to explain the NIR excess in SC-SN 2012 due to the substantial extinction by the CS dust. This highlights the importance of detailed radiation transfer simulations.
\begin{figure}
  \includegraphics[width=\columnwidth]{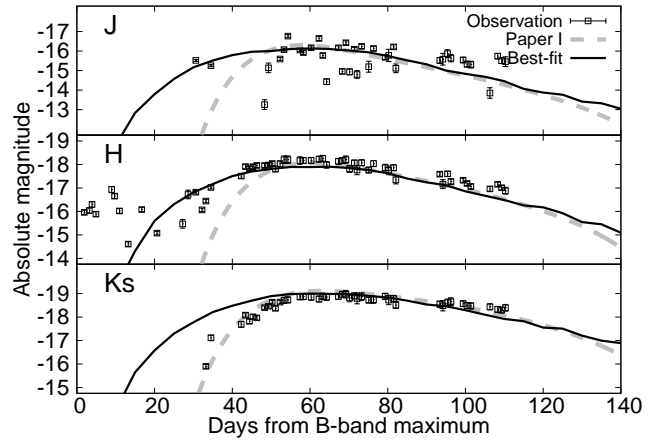}
  \caption{$J$-, $H$-, and $K_s$-band light curves in the updated best-fit model in Table 2 (black solid lines). The gray dashed lines show the original light curves in the disk model of Paper I calculated using the simple method. The light curves of the NIR excess seen in SC-SN 2012dn are also shown (black squares).}
\end{figure}

To evaluate the effects of the extinction, the following values ($\tau_{\parallel}, \tau_{\perp}$) are calculated for the former best-fit models in Paper I (see Table 1).
\begin{eqnarray}
\tau_{\parallel} (\nu) &\equiv& \int_{r_{\mathrm{in}}}^{r_{\mathrm{out}}} \kappa_{\mathrm{ext}} (\nu) \rho_{\mathrm{dust}} (r) dr\\
&=& \kappa_{\mathrm{ext}} (\nu) \rho_{\mathrm{dust}} (r_{\mathrm{in}}) \frac{r_{\mathrm{in}}}{r_{\mathrm{out}}} (r_{\mathrm{out}} - r_{\mathrm{in}})
\end{eqnarray}

\begin{eqnarray}
\tau_{\perp} (\nu) &\equiv&  \int_{0}^{r_{\mathrm{in}} \tan(\theta_0/2)} \kappa_{\mathrm{ext}} (\nu) \rho_{\mathrm{dust}} (r_{\mathrm{in}}) dr\\
&=& \kappa_{\mathrm{ext}} (\nu) \rho_{\mathrm{dust}} (r_{\mathrm{in}}) r_{\mathrm{in}} \tan \biggl(\frac{\theta_0}{2} \biggr)\\
&=& \frac{r_{\mathrm{out}} \tan \bigl(\frac{\theta_0}{2} \bigr)}{r_{\mathrm{out}} - r_{\mathrm{in}}} \tau_{\parallel} (\nu)
\end{eqnarray}

The first value, $\tau_{\parallel}$, expresses typical optical depth along the disk plane (for the disk model) or that along the jet axis (for the jet model). The second value, $\tau_{\perp}$, expresses typical optical depth perpendicular to the disk plane (for the disk model) or that perpendicular to the jet axis (for the jet model). In the previous best-fit jet model, the optical depth $\tau_{\parallel}$ is indeed much larger than unity (see Table 1). Therefore, the NIR echo flux does not become larger, even if the mass of the CS dust is increased. We conclude that the jet-like geometry cannot explain the observed NIR excess in SC-SN 2012dn. 

Finally, the validity of the value of $\theta_{\rm{obs}}$ in the best-fit model in Paper I is checked here. Figure 3 shows the light curves of the NIR echoes for the updated disk model with various values of $\theta_{\mathrm{obs}}$, where only the value of $\theta_{\rm{obs}}$ is changed from the new-best fit value. From this result, we conclude that the value of $\theta_{\mathrm{obs}}$ should be less than 15 degree.
\begin{figure}
  \includegraphics[width=\columnwidth]{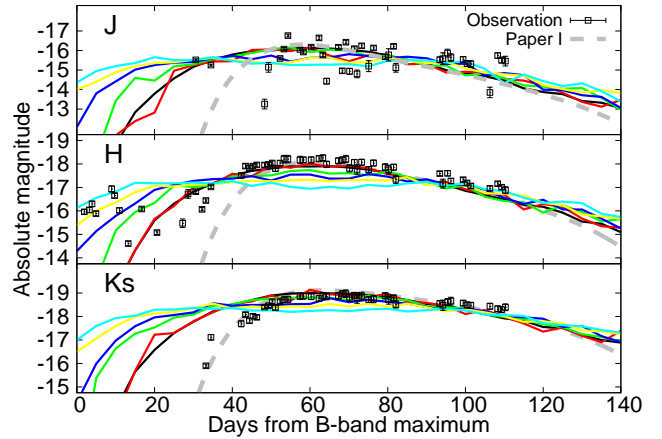}
  \caption{Same as Figure 2, but for the updated best-fit disk model with different values of $\theta_{\rm{obs}}$ (black solid line for $\theta_{\rm{obs}} = 5$ degree, red solid line for $\theta_{\rm{obs}} = 15$ degree, green solid line for $\theta_{\rm{obs}} = 25$ degree, blue solid line for $\theta_{\rm{obs}} = 35$ degree, yellow solid line for $\theta_{\rm{obs}} = 45$ degree, light blue solid line for $\theta_{\rm{obs}} = 55$ degree).}
\end{figure}

\subsection{Predicted properties of the polarized-scattered echoes}
With the detailed transfer simulations, we address polarization signatures to distinguish the nature of the CS dust. If the (new) best-fit CS dusty disk derived from the observed NIR excess is viewed exactly from the face-on direction, no net polarization is expected due to the projected circular symmetry of CS dust as viewed from the observer. Even in the case with $\theta_{\rm{obs}}=15$ degree, which is the maximum acceptable value for SC-SN 2012dn, $P \lesssim 1$ \% (see below). Unfortunately, there is no published polarimetric observation for SC-SN 2012dn so far.

However, if the dusty disk derived for SC-SN 2012dn is a common feature shared by (a part of) SC-SNe, such systems should be observed at various viewing angles, once a large sample of SC-SNe is obtained. To predict observables for such future observations, the polarization degree of SNe Ia in the $B$ band with the updated best-fit CS dusty disk (see Table 2) are calculated for various values of $\theta_{\mathrm{obs}}$, as shown in Figure 4a. In the case with $\theta_{\mathrm{obs}} = 10$ degree (consistent with the situation for SC-SN 2012dn), we predict low polarization (always $\lesssim 1$\%) because of the projected circular symmetry of CS dust as viewed from an observer. For larger $\theta_{\mathrm{obs}}$, the polarization degree is higher, where the peak of the polarization is reached around $\sim 60$ days since the $B$-band maximum. This coincides with the peak of the NIR excess, since both the scattering and thermal echoes come from the same CS dust. The polarization degree is maximized at $\theta_{\mathrm{obs}} \sim 90$ degree. Figure 4b shows the time evolution of the position angle of the polarization for each viewing angle shown in Figure 4a. As clearly shown for $\theta_{\mathrm{obs}} = 30$ and $50$ degree, the position angle evolves from 90 to 0 (180) degree at a certain time when the polarization degree is quickly increased. 
\begin{figure*}
  \includegraphics[width=\columnwidth]{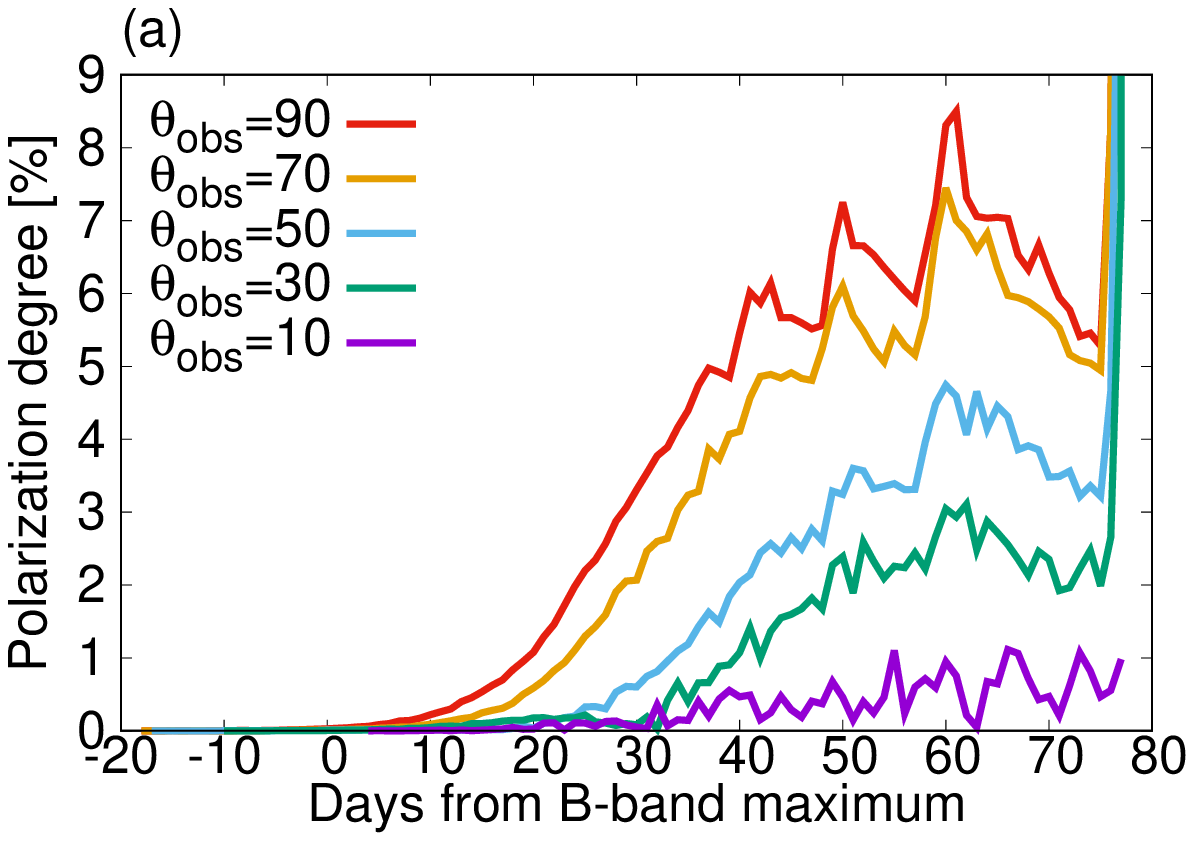}
  \includegraphics[width=\columnwidth]{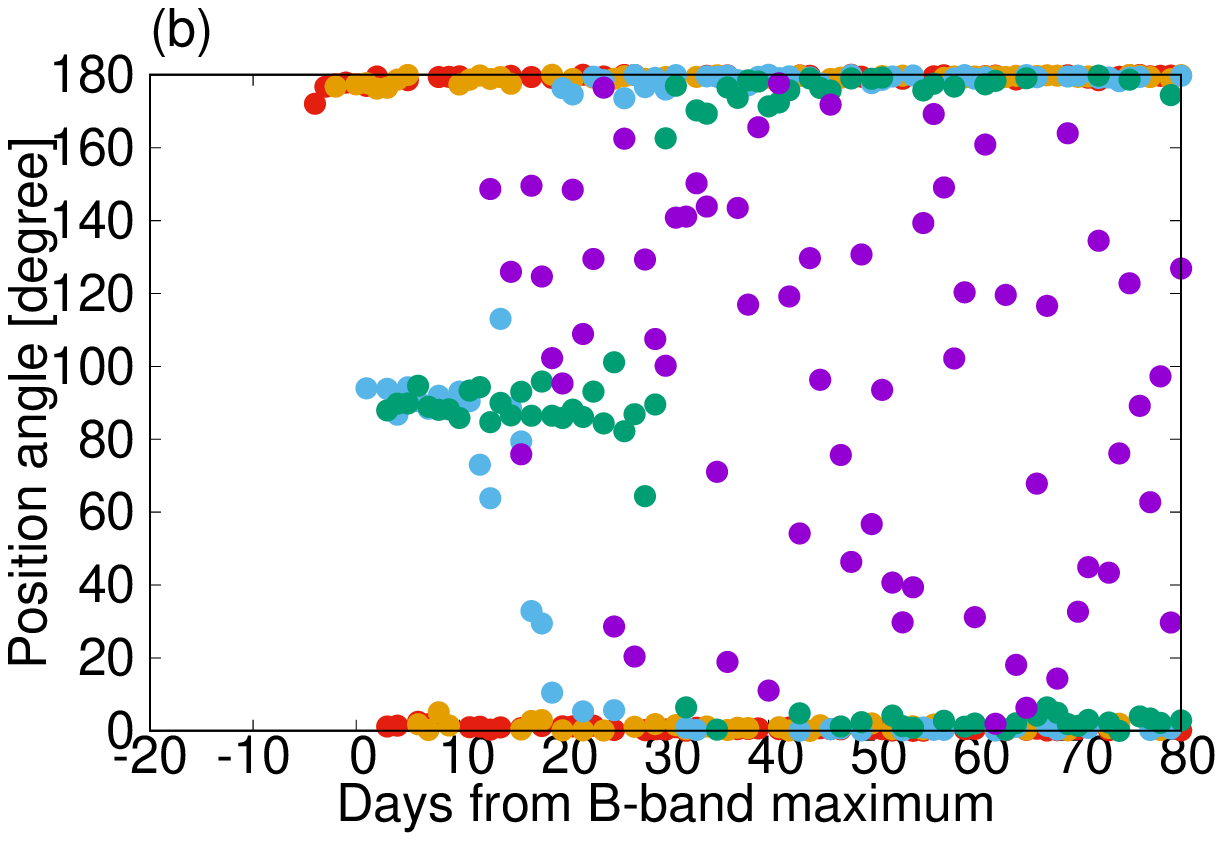}
  \caption{(a) Time evolution of the polarization degree in the $B$ band for the updated best-fit disk model with various values of $\theta_{\rm{obs}}$ (violet for $\theta_{\rm{obs}} = 10$ degree, green for $\theta_{\rm{obs}} = 30$ degree, light blue for $\theta_{\rm{obs}} = 50$ degree, orange for $\theta_{\rm{obs}} = 70$ degree, red for $\theta_{\rm{obs}} = 90$ degree). (b) Time evolution of the position angle of the polarization shown in (a).}
\end{figure*}

The physical origin of the general behavior as described above has been explained for a different model setup by \citet{Nagao2017b}. To understand these features more deeply, the position angles at each spatial position of the CS dusty disk are shown in Figure 5, though our targets are practically point sources. The position angle of the scattered echo depends on a position of the disk from which the echo is originated. The photons from the closest/farthest side of the disk (the first/third component) have the position angle that is parallel to the $y$ axis (90 degree) for the observer with ($r$,$\theta$,$\phi$)=($\infty, \theta_{\mathrm{obs}},0$) in the spherical coordinates, while the other photons (the second component) have the vertical component (0 (180) degree) \citep[see Figure 8 in][]{Nagao2017b}. Therefore, the position angle of the net polarization of the echo temporally evolves in the disk model, following the change of the dominating component from the first to second. The polarization degree is typically lower during the time when the first component dominates, because the flux of the first component is smaller than that of the second component. After then, the polarization degree is determined by the relative flux of each component. This simple interpretation applies when $\tau_{\parallel}$ (optical) $\lesssim 1$, i.e., when the effects of multiple scattering for polarization are not important as in the best-fit case. The relative flux and delay time of each component depend on $\theta_{\rm{obs}}$. For larger $\theta_{\rm{obs}}$, the duration when the first component dominates is shorter due to the smaller area where the first component is originated. Indeed, the first component is created by being scattered with smaller scattering angle, which leads to smaller polarization degree. Thus the second component starts dominating earlier, and also at higher flux \citep[see \S 3.1.2 in][]{Nagao2017b}. Especially in the case with $65 \gtrsim \theta_{\rm{obs}} \gtrsim 115$ degree ($\sim 40$ \% over the entire solid angle), where the CS disk is between an SN and an observer, the intrinsic SN light is also extinguished by the CS disk. Since the optical depth along the disk ($\tau_{\parallel}$) is 0.536, the SN become fainter by $\sim 0.6$ magnitude. Therefore, the contribution of the polarized echo component is more enhanced. Finally, when the third component reach to the observer, the second component has already been enhanced. Therefore, the third component basically do not play any important role for the net SN polarization.
\begin{figure*}
  \includegraphics[width=\columnwidth]{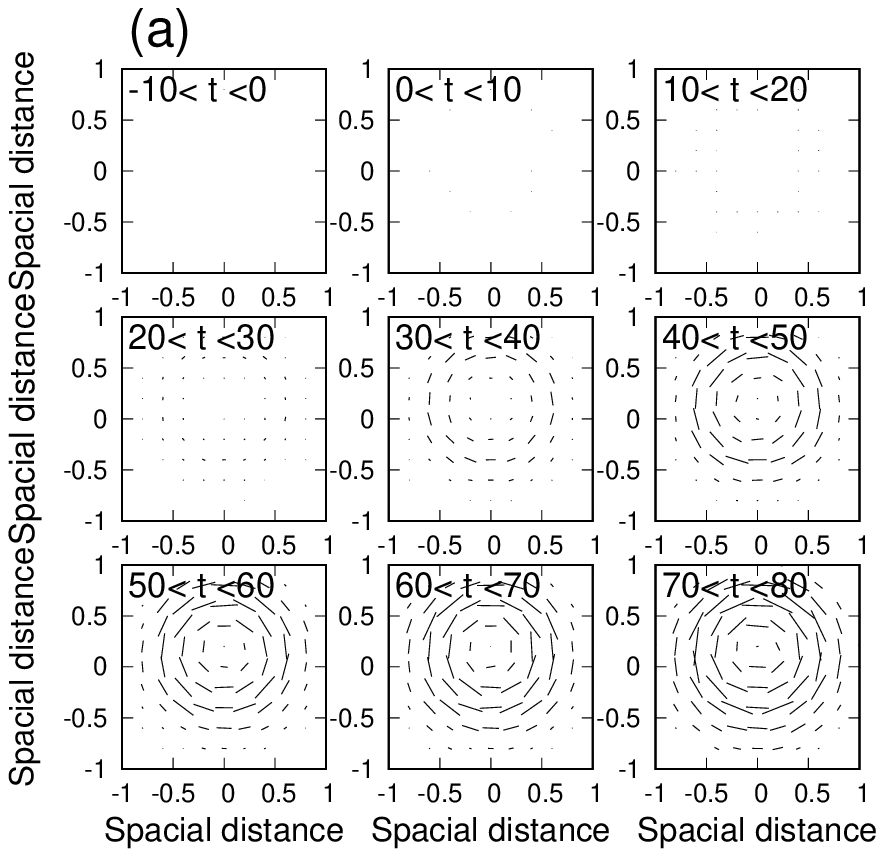}
  \includegraphics[width=\columnwidth]{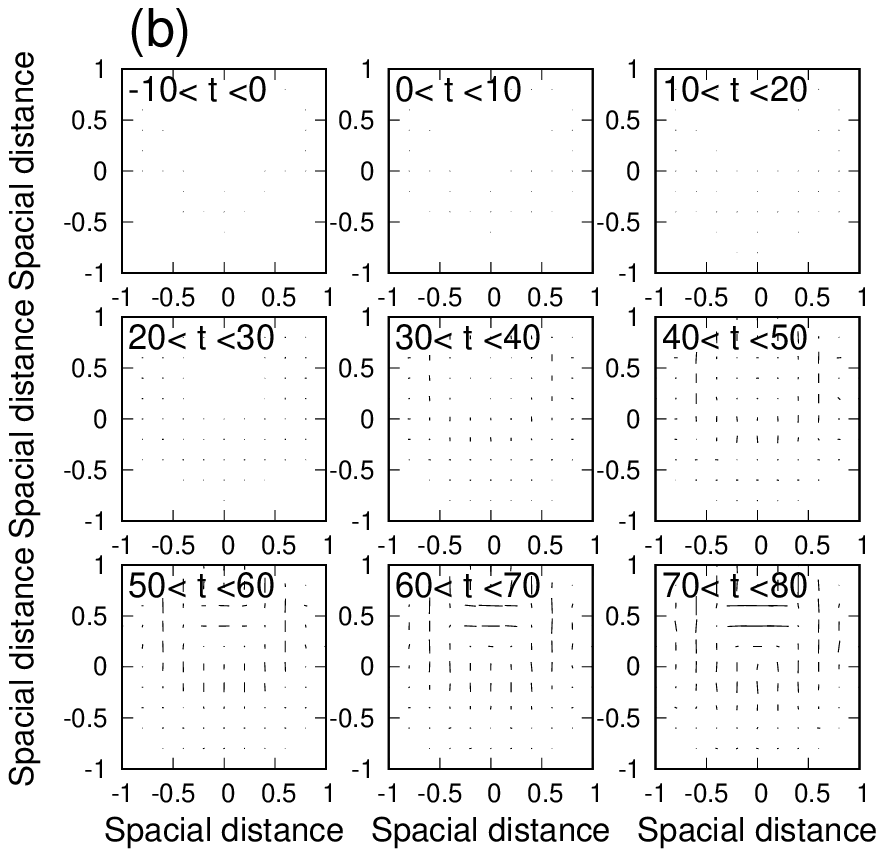}
  \includegraphics[width=\columnwidth]{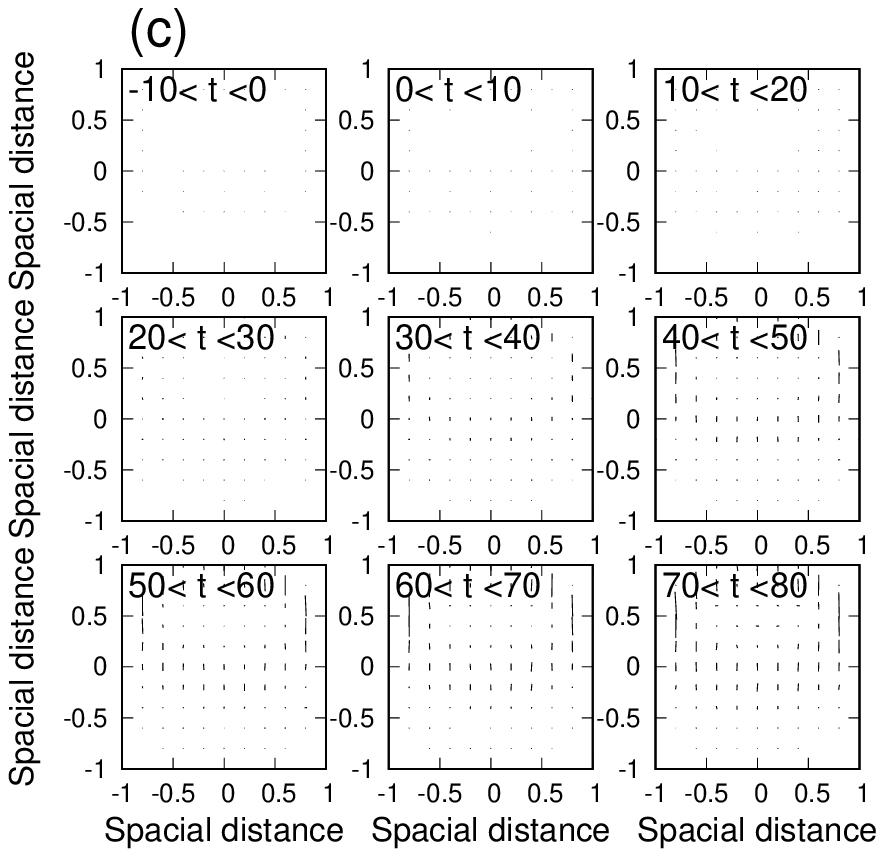}
  \includegraphics[width=\columnwidth]{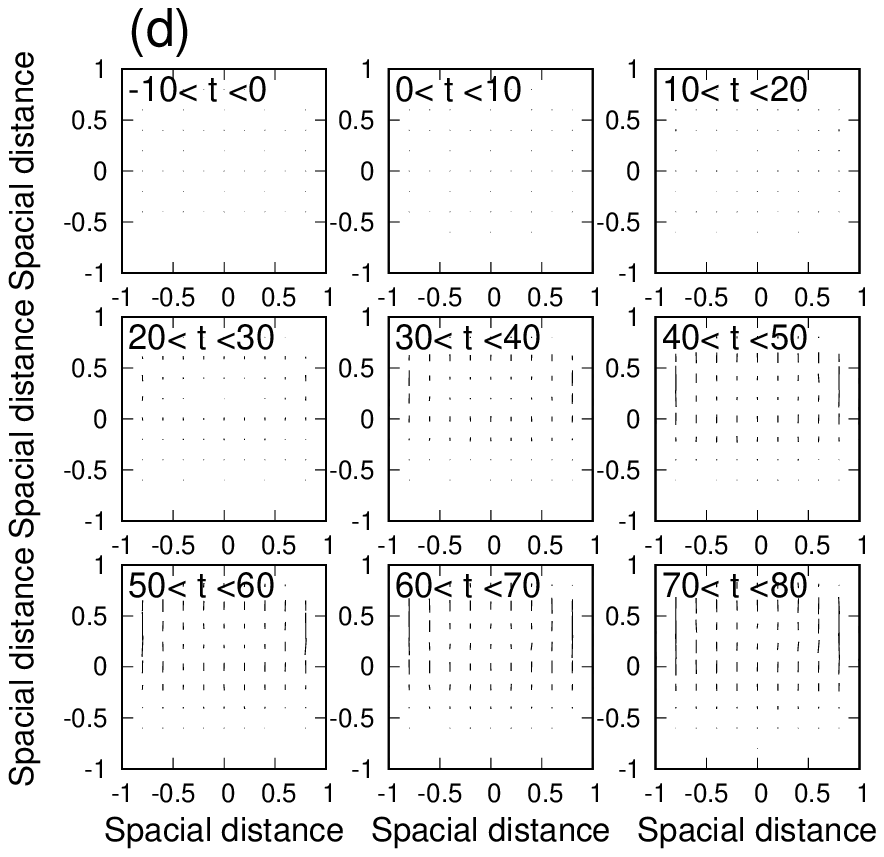}
  \includegraphics[width=\columnwidth]{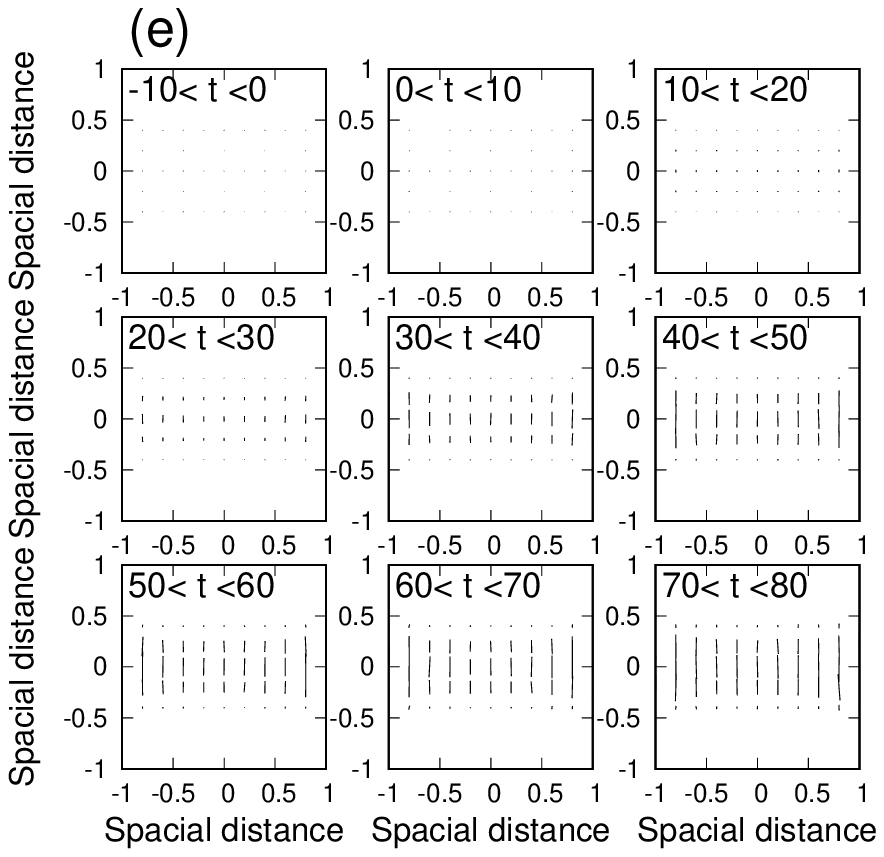}
  \caption{Polarization map of the scattered echo in the $B$ band for the updated best-fit disk model with various values of $\theta_{\rm{obs}}$ ((a) for $\theta_{\rm{obs}} = 10$ degree, (b) for $\theta_{\rm{obs}} = 30$ degree, (c) for $\theta_{\rm{obs}} = 50$ degree, (d) for $\theta_{\rm{obs}} = 70$ degree, (e) for $\theta_{\rm{obs}} = 90$ degree). Each panel shows a polarization map at each time ([day]) from $B$-band maximum. The direction along the y axis corresponds to 0 (180) degree for position angle, while the direction along the x axis corresponds to 90 degree for position angle.}
\end{figure*}

The polarization degree is sensitive to the optical depth of the CS dust in a different manner from the NIR echo. Figure 6 shows the time evolution of the polarization degree in the $B$-band for the best-fit CS disk in Table 2, except for the mass, which is varied as experiment. As the total disk mass is set higher, the polarization degree becomes higher. However, if the total mass becomes about two times higher than the best-fit value ($\tau_{\parallel}$ (opt) $\gtrsim 1$), the polarization degree does not become higher anymore, because of the effects of the multiple scattering. Once multiple scattering contributes to the scattering processes, the anisotropically distributed polarization angle due to the asphericity in the distribution of the CS dust is randomized, leading to cancellation of the polarization.
\begin{figure*}
  \includegraphics[width=\columnwidth]{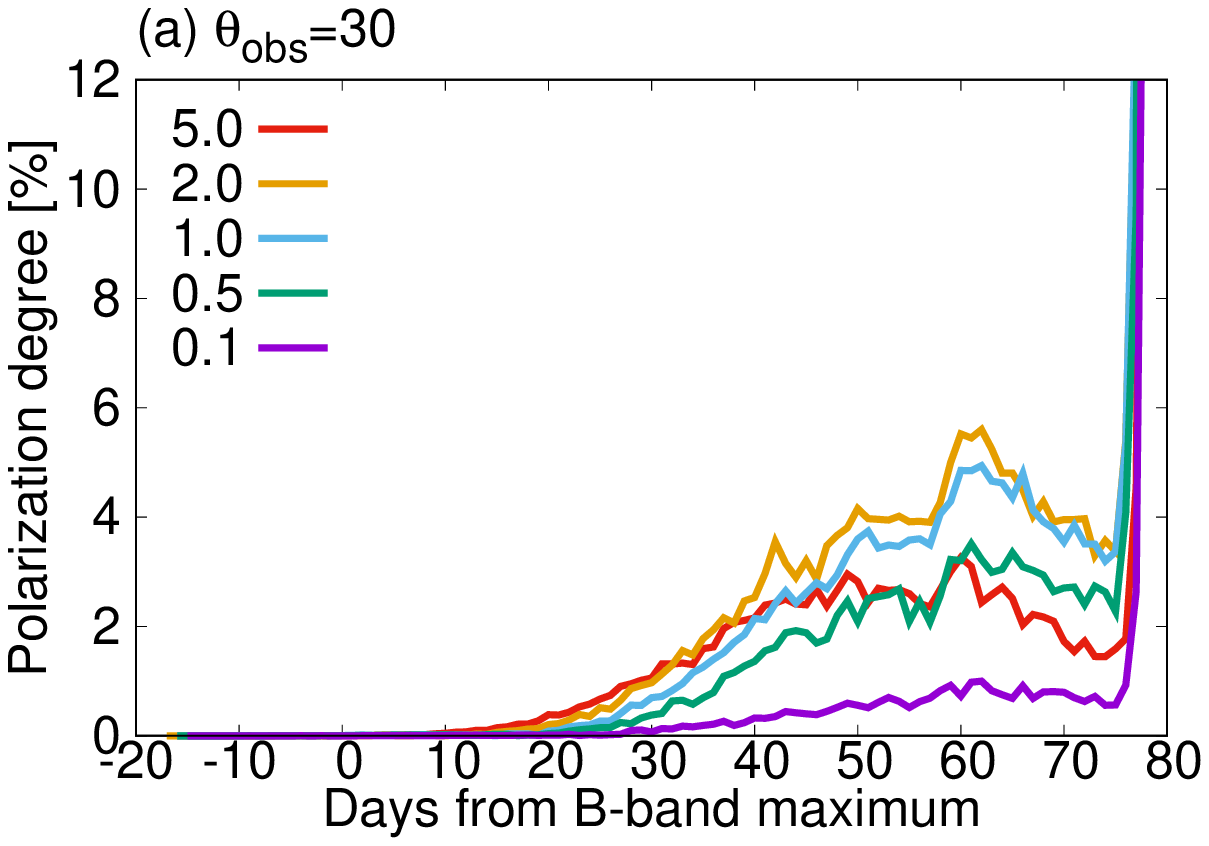}
  \includegraphics[width=\columnwidth]{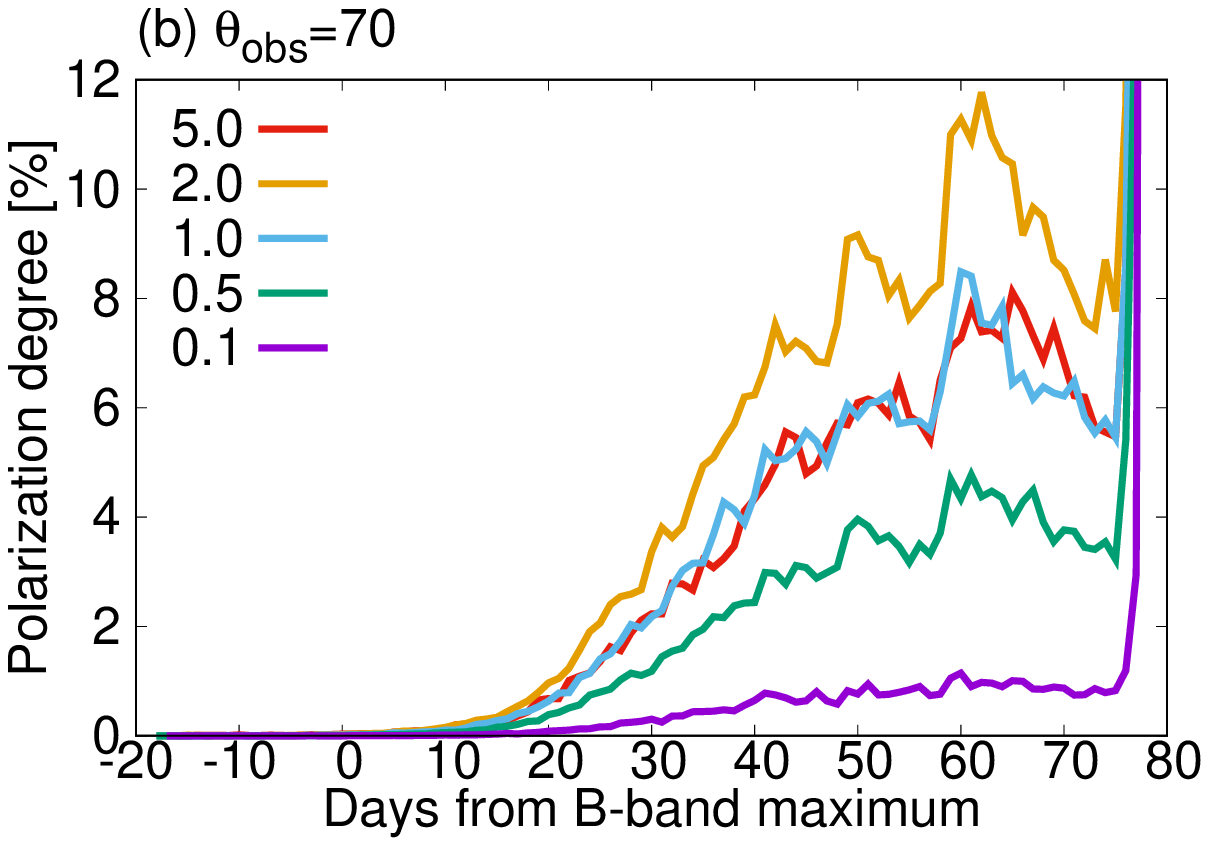}
  \caption{Same as Figure 4(a), but for the updated best-fit disk model with various values of the total dust mass, where (a) $\theta_{\rm{obs}} = 30$ and (b) $\theta_{\rm{obs}} = 70$ degree (violet for 0.1 times, green for 0.5 times, light blue for 1.0 times, orange for 2.0 times, red for 5.0 times the CS dust mass derived for SC-SN 2012dn).}
\end{figure*}

Since the optical depth of the CS dusty disk is higher for the SN light at shorter wavelength, it is expected to observe different values of the polarization degree for different wavelength. Figure 7 shows the time evolution of the polarization degree in different wavelength for the best-fit CS disk. In the best-fit disk, $\tau_{\parallel} \lesssim 1$ even for the $U$-band light. Thus, the value of the polarization degree is maximized at the $U$ band. The expected wavelength dependence of the polarization curve roughly follows that of scattering coefficient of CS dust ($\kappa_{\rm{scat}} \propto \lambda^{-1}$ in our dust model), and the peak wavelength in the polarization curve is reached at a wavelength where the optical depth along the disk is $\sim 2$.
\begin{figure*}
  \includegraphics[width=\columnwidth]{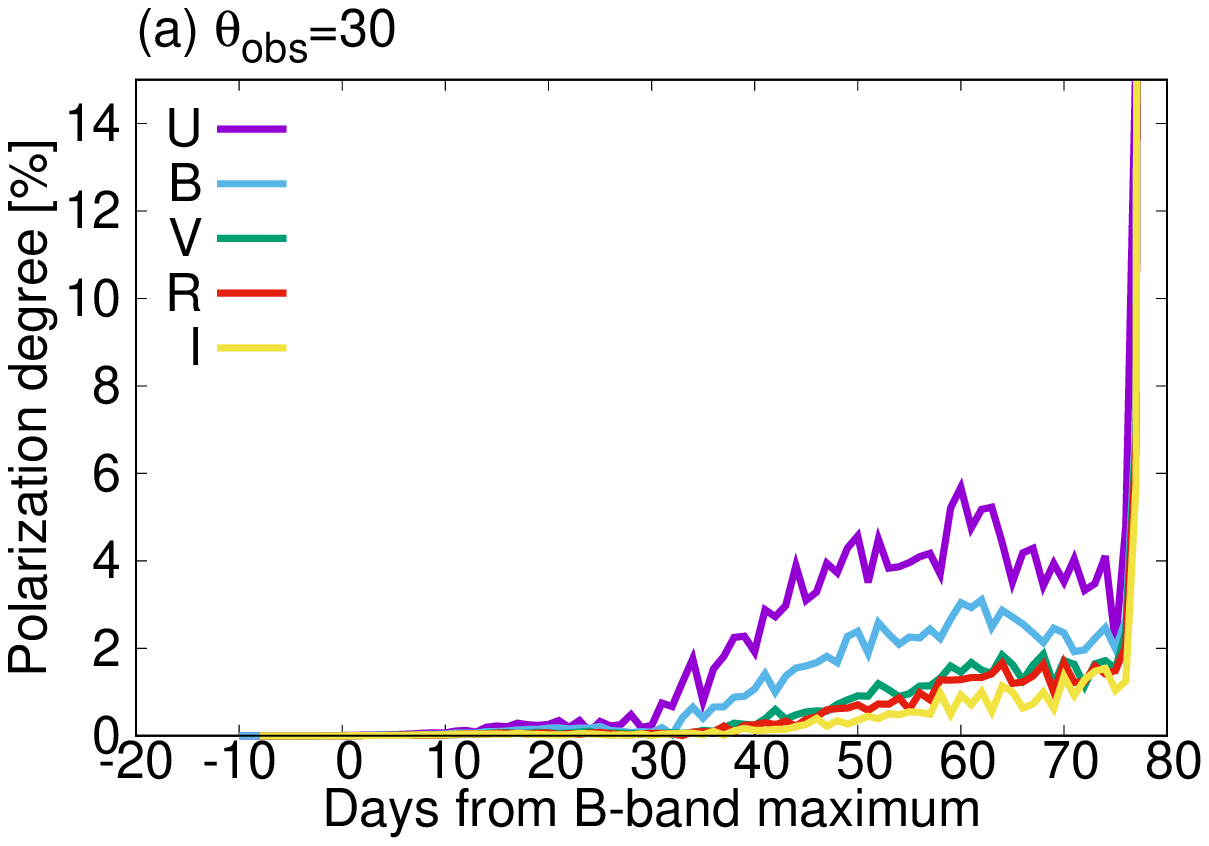}
  \includegraphics[width=\columnwidth]{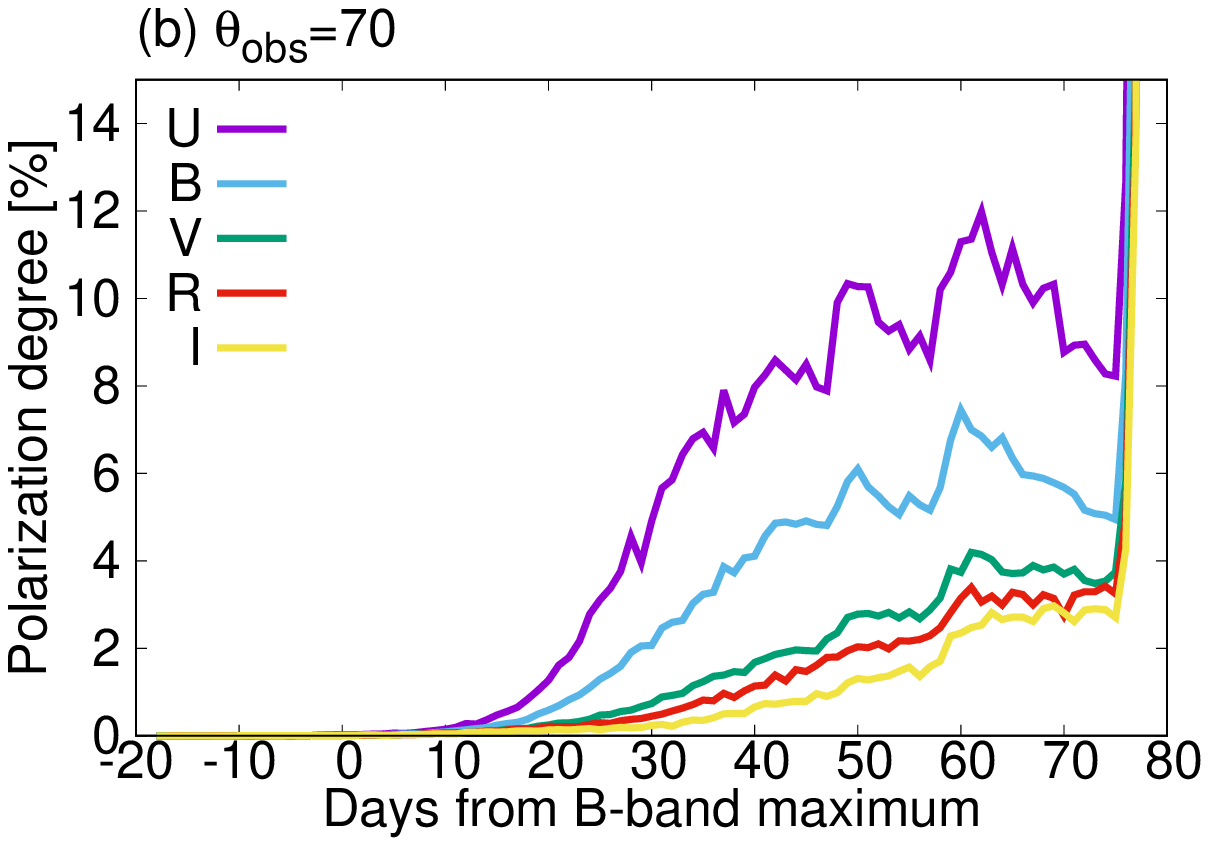}
  \caption{Same as Figure 4(a), but at various values of wavelength, where (a) $\theta_{\rm{obs}} = 30$ and (b) $\theta_{\rm{obs}} = 70$ degree (violet for the $U$ band, light blue for the $B$ band, green for the $V$ band, red for the $R$ band, yellow for the $I$ band).}
\end{figure*}

The polarization degree is also sensitive to the geometry of the CS dust in a different manner from the NIR echo. Unlike the NIR echo, polarization from symmetrically-distributed CS dust is totally cancelled out. In the case with SC-SN 2012dn, the jet model could be rejected even from the NIR echo alone. In other cases, polarimetric observations are important to clarify the amount and distribution of CS dust around SNe. For example, in the case where the same parameters with the best-fit CS dust model are adopted but with the lower total dust mass ($\tau_{\parallel}$ (opt) $\lesssim 1$), the expected NIR echoes from the disk and jet models are essentially the same, but the expected polarization degree can be totally different.

\section{Discussion and conclusions}
We have investigated the detailed distribution of the CS dust to account for the observed NIR excess in SC-SN 2012dn using the 3D radiation transfer code. We further presented predictions on the time evolution of the polarization with the derived CS dusty disk. Here, discussions are given for observability of the polarization, and also for implications for CS dust around SC-SNe and normal SNe Ia.

If the dusty CS disk around SC-SN 2012dn is a common feature shared by the SN2012dn-like SC-SNe, such systems will be observed for other SC-SNe from various viewing directions, once a large sample of SC-SNe is obtained. Indeed, SC-SNe may be divided into two classes according to the small sample analyzed so far; SN 2009dc-like SC-SNe are characterized by large luminosities, while SN 2012dn-like SC-SNe show luminosities similar to normal SNe Ia \citep[e.g.,][]{Chakradhari2014}. Spectral features may also follow this classification scheme \citep[e.g.,][]{Chakradhari2014}. Under this hypothesis, we estimate their frequency and discuss a strategy for the proposed polarimetric observations to probe the dusty CS environment. The frequency of SN 2012dn-like SC-SNe within the distance of SC-SN 2012dn ($\sim 40$ Mpc) is $\sim 0.04$ per year, assuming that the frequency is roughly $\sim 0.5$ \% of that of normal SNe Ia. Since the peak apparent magnitude of SC-SN 2012dn is $\sim 14$ magnitude, most of them are easily discovered by recent large optical SN survey networks. The apparent magnitude of SN 2012dn is $\sim 17$ magnitude at $\sim 60$ days after the $B$-band peak, when the expected polarization degree is maximized ($\gtrsim 4$ \% for $\theta_{\rm{obs}} \gtrsim 60$ degree). This polarization degree can be observed by one-meter telescopes in the imaging polarimeter mode, though the expected frequency of such an event is as low as one in $\sim 20$ years. Telescopes with diameter of 4-8 m can detect a few \% level of polarization from objects whose apparent magnitude is $\sim 19.5$ magnitude, thus this can be studied for one or a few events per year. Such SNe reach $\sim 16.5$ magnitude at the peak, thus this is not limited by the discovery capability.

In Paper I, we assumed that the light curves of SC-SN 2009dc purely consisted of an SN light without an echo. In addition, we derived the conservative upper limit of the mass of CS dust around SC-SN 2009dc, by the requirement that the predicted echo luminosity should not exceed the observed one at any bands and epochs. The estimated upper limit was $\sim 1.9 \times 10^{-6} \; M_{\odot} \rm{yr}^{-1}$. Here, we again derive the upper limit, using not NIR observations but optical polarimetric observations. The requirement in this case is the same with that in Paper I: the predicted polarization degree should not exceed the observed one at any epochs. We assume that SC-SN 2009dc has an identical CS disk to that around SC-SN 2012dn, except for the CS dust mass. Here, we assume that the value of $\theta_{\mathrm{obs}}$ is $\sim 60$ degree, which is the most likely value for a randomly distributed viewing angle. \citet{Tanaka2010} has reported that the continuum polarizations of SC-SN 2009dc at 5.6 and 89.5 days since the $B$-band maximum are small ($<$ 0.3 \%). With these constraints, we derive the upper limit on the CS dust mass as $\sim 0.06$ times the CS dust mass derived for SC-SN 2012dn. The corresponding (gas) mass-loss rate for this CS dust is $\sim 9.6 \times 10^{-7} \; M_{\odot} \rm{yr}^{-1}$, which is slightly a deeper limit than that derived in Paper I. In this case, the polarimetric observation could be even more useful to constraint the CS dust mass than the NIR observation. This comes mainly from the fact that the contaminating SN component is virtually negligible in the polarization signal, highlighting the power of the polarization diagnostics. In short, we reconfirm that SC-SN 2009dc is in a clean environment, though we cannot reject a possibility that SC-SN 2009dc has a symmetric CS dust as viewed from the observer.

Recently, the origin of extinction toward highly reddened SNe Ia ($E(B-V) \sim 1$ mag) has been debated, which may provide a hint to understand the progenitors of SNe Ia \citep[e.g.,][]{Amanullah2015}. The SNe tend to show unusual extinction curves (low $R_{V}$). This can be explained either by an enhanced abundance of small dust grains in the interstellar scale \citep[e.g.,][]{Gao2015} or by multiple scattering effects within CS dust in the circumstellar scale \citep[e.g.,][]{Wang2005, Patat2005, Goobar2008}. The present study on the polarization signature from the CS dust has implications for this issue as well. If we consider a case that the origin of the extinction is CS dust, and assume the same disk-like structure for the CS dust as is derived for SC-SN 2012dn, then a fraction of SNe Ia must be observed from the edge-on direction. Since the value of $E(B-V) \sim 1$ mag corresponds to an optical depth $\tau \sim 4$ for both the LMC and MW dust \citep{Nagao2016}, and the expected polarization degree is $\sim 10$ \% if viewed from such a direction. In other words, the observed small polarization degree already places some constraints on the CS multiple scattering scenario as follows. The CS dust must be distributed more or less in a spherical manner if the observed extinction is associated with the CS environment. A disk-like CS environment as is derived for SC-SN 2012dn, which might be expected from (at least a branch of) the SD scenario \citep[][]{Booth2016}, would not account for the extinction without conflicting to the polarization constraint.

The nature of progenitors of SNe Ia is still unclear \citep[see, e.g.,][for a review]{Maeda2016b}. There are three main scenarios for the progenitors: the SD, DD and CD scenarios (see \S 1). The progenitor systems in the SD scenario are expected to eject a certain amount of CS medium ($\dot{M} \sim 10^{-7}$-$10^{-6} \; M_{\odot}$ yr$^{-1}$) to realize stable mass transfer to the primary WDs \citep[][]{Nomoto1982}. Here, we consider a case that this CS medium is aspherical as the dusty disk derived for SC-SN 2012dn, and viewed from an SN through the CS medium from the edge-on direction. Such a situation is expected in (at least a branch of) the SD scenario \citep[e.g.,][]{Booth2016}. From our results in Section 4.2, $\sim 1$ \% and $\sim 0.1$ \% of polarization degree are expected for cases with $\dot{M} \sim 10^{-6}$ and $10^{-7} \; M_{\odot}$ yr$^{-1}$, respectively. Once a large sample of SNe Ia is constructed, existence of such SNe with relatively high polarization degree is thus predicted for the SD scenario. The polarimetric observation is thus a powerful and unique tool to investigate distributions of CS medium around not only SC-SNe Ia but also normal SNe Ia. We also note that there is a class of SNe Ia associated with a dense CSM \citep[e.g.,][]{Hamuy2003}, and the polarimetric observations are highly encouraged for this class as well.

Our conclusions are summarized as follows:
\begin{enumerate}
\item The geometry of the CS dust around SC-SN 2012dn must be a disk-like. 
\item The derived mass loss rate is high, $\sim 1.6 \times 10^{-5} \; M_{\odot}$ yr$^{-1}$ for the mass loss velocity of $\sim 10$ km s$^{-1}$, which could be consistent with a branch of the SD model associated with the high mass transfer rate. 
\item While no polarimetric observations are available in the literature for SC-SN 2012dn, we predict a very low polarization level for this SN because of the viewing direction aligned to the polar direction.
\item If the same CSM is viewed from different directions, the polarization degree can reach to $\sim 8\%$. 
\item The polarization mentioned above is predicted to follow a low degree of polarization, and then a change in the position angle by 90 degree when the polarization increases. 
\item The signal should be easily distinguishable from the background interstellar polarization and intrinsic SN polarization in the temporal and wavelength dependence (among other unique features). 
\item If the CS environment found for SC-SN 2012dn is shared by (a part of) SC-SNe Ia, a good fraction of SC-SNe Ia should show large polarization.
\item The polarimetric observations thus provide a powerful diagnostic to investigate the existence and nature of CS dust around SNe Ia. Once SN 2012dn-like SC-SNe are discovered, multi-band polarimetric observations especially around $\sim 60$ days after the B-band maximum are important for revealing the geometry of the CS dusty disk around the SNe. 
\item The estimated frequency of SN 2012dn-like SC-SNe whose polarization can be observed by 4-8m class telescopes is one or a few events per year.
\item The derived upper limit of the mass-loss rate corresponding to the mass of CS dust around another SC-SN 2009dc is $\sim 9.6 \times 10^{-7} \; M_{\odot}$ yr$^{-1}$, assuming that SC-SN 2009dc has an identical CS disk to that around SC-SN 2012dn and $\theta_{\rm{obs}} \sim 60$ degree.
\item The CS dust around normal SNe Ia must be distributed more or less in a spherical manner if the observed extinction toward highly reddened SNe Ia is associated with the CS environment.
\item The geometry of CS dust around normal SNe Ia, which is expected in the SD scenario, can be investigated by deep polarimetric observations for a sample of SNe Ia.
\end{enumerate}

\section*{Acknowledgements}
The authors thank Francisco F\"{o}rster for giving us constructive comments at the YITP workshop YITP-T-16-05 on `Transient Universe in the Big Survey Era: Understanding the Nature of Astrophysical Explosive Phenomena'. The authors thank the Yukawa Institute for Theoretical Physics at Kyoto University, where the workshop was held. Simulations were in part carried out on the PC cluster at Center for Computational Astrophysics, National Astronomical Observatory of Japan. This work was partially supported by the Optical and Near-infrared Astronomy Inter-University Cooperation Program. The work has been supported by Japan Society for the Promotion of Science (JSPS) KAKENHI Grant 17J06373 (T.N.), 17H02864 (K.M.), and 17K14253 (M.Y.).

%%%%%%%%%%%%%%%%%%%%%%%%%%%%%%%%%%%%%%%%%%%%%%%%%%



%\bibliographystyle{mnras}


%%%%%%%%%%%%%%%%%%%%%%%%%%%%%%%%%%%%%%%%%%%%%%%%%%

%%%%%%%%%%%%%%%%% APPENDICES %%%%%%%%%%%%%%%%%%%%%

%\appendix

%%%%%%%%%%%%%%%%%%%%%%%%%%%%%%%%%%%%%%%%%%%%%%%%%%

% Don't change these lines
\bsp	% typesetting comment
\label{lastpage}
\end{document}